\documentclass{aa}  

\usepackage{booktabs}

\usepackage{graphicx}
\usepackage{natbib}
\usepackage{txfonts}
\usepackage{threeparttable}
\usepackage{siunitx}
\usepackage{color}
\usepackage{xcolor}
\definecolor{mhi}{rgb}{0.6,0,0.6}

\definecolor{msp}{rgb}{0.0,0.6,0.6}

\begin{document}

   \title{The Cen\,A galaxy group: dynamical mass and missing baryons\thanks{Based on observations collected at the European Organisation for Astronomical Research in the Southern Hemisphere under the ESO program 0101.A-0193.}} 
   
\titlerunning{The dynamical mass of Cen\,A}
   \author{
          Oliver M\"uller\inst{1,}\inst{2}
                              \and
        Federico Lelli\inst{3}
        \and
        Benoit Famaey\inst{2}
        \and
        Marcel S. Pawlowski\inst{4}
                 \and
         Katja Fahrion\inst{5,}\inst{6}
         \and
         Marina Rejkuba\inst{6}
          \and
         Michael Hilker\inst{6}
          \and
         Helmut Jerjen \inst{7}
          }
          
           \institute{Institute of Physics, Laboratory of Astrophysics, École Polytechnique Fédérale de Lausanne (EPFL), 1290 Sauverny, Switzerland\\
 \email{oliver@oliver-mueller.ch}
\and
Universit\'e de Strasbourg, CNRS, Observatoire astronomique de Strasbourg, UMR 7550, F-67000 Strasbourg, France
 \and
INAF, Arcetri Astrophysical Observatory, Largo Enrico Fermi 5, I-50125, Florence, Italy
 \and
Leibniz-Institut fur Astrophysik Potsdam (AIP), An der Sternwarte 16, D-14482 Potsdam, Germany
\and
 European Space Agency (ESA), European Space Research and Technology Centre (ESTEC), Keplerlaan 1, 2201 AZ Noordwijk, The Netherlands
\and
 European Southern Observatory, Karl-Schwarzschild Strasse 2, 85748, Garching, Germany
\and
 Research School of Astronomy and Astrophysics, Australian National University, Canberra, ACT 2611, Australia
}

   \date{Received September 15, 1996; accepted March 16, 1997}

   \abstract{
   The nearby elliptical galaxy Cen\,A is surrounded by a flattened system of dwarf satellite galaxies with coherent motions. Using a novel Bayesian approach, we measure the mean rotation velocity $v_{\rm rot}$ and velocity dispersion $\sigma_{\rm int}$ of the satellite system. We find $v_{\rm rot}/\sigma_{\rm int} \simeq 0.7$ indicating that the satellite system has non-negligible rotational support. Using Jeans' equations, we measure a circular velocity of 258 km\,s$^{-1}$ and a dynamical mass of $1.2\times 10^{13}$ M$_\odot$ within 800\,kpc. In a $\Lambda$CDM cosmological context, we find that the Cen\,A group has a baryon fraction $M_{\rm b}/M_{200}\simeq0.035$ and is missing $\sim$77$\%$ of the cosmologically available baryons. Consequently, Cen\,A should have a hot intergalactic medium with a mass of $\sim$8$\times$10$^{11}$ M$_\odot$, which is more than $\sim$20 times larger than current X-ray estimates. Intriguingly, The whole Cen\,A group lies on the baryonic Tully-Fisher relation defined by individual rotationally supported galaxies, as expected in Milgromian dynamics (MOND) with no need of missing baryons.
   }

   \maketitle
%

\section{Introduction}

Galaxy groups are a key testbed for the $\Lambda$ Cold Dark Matter ($\Lambda$CDM) cosmological model \citep{2010A&A...523A..32K,2017ARA&A..55..343B,Oppenheimer2021} as well as for alternative theories  \citep{2018MNRAS.473.4033B,Milgrom2019}. In the Local Volume ($D<11$ Mpc), about half of all major galaxies reside in virialized groups, while the remaining half constitutes the so-called ``field'' population \citep{Karachentsev2005}. More massive galaxy clusters, which are absent in the Local Volume, are estimated to contain a minor fraction of the galaxy population, about 10$\%-$15$\%$ \citep{Karachentsev2005}. Unfortunately, estimating the dynamical mass ($M_{\rm dyn}$) of galaxy groups is more challenging than for galaxy clusters. 
Only a sub-sample of galaxy groups have a high-density hot medium that can be studied with existing X-ray telescopes to estimate $M_{\rm dyn}$ from hydrostatic equilibrium (e.g. see \citealt{2017ApJ...843...16K} and references therein). Moreover, galaxy groups are too diffuse to produce detectable gravitational lensing signal from background galaxies. The only remaining approach is using the line-of-sight velocities of galaxy members, as it has been pioneered by \citet{1933AcHPh...6..110Z} almost a century ago.

Dynamical mass estimates for galaxy groups usually rely on the Virial theorem and/or on the ``zero-velocity surface'' method \citep{1981Obs...101..111L,1986ApJ...307....1S,Karachentsev2005,2009MNRAS.393.1265K,2015AJ....149...54T,2018A&A...609A..11K}. Both methods assume that the member galaxies are isotropically distributed (spherical symmetry) and follow random orbits. These assumptions appear unreasonable for our own Local Group: most dwarf satellites of the Milky Way and M31 are distributed in narrow planar structures with significant angular momentum \citep{2012MNRAS.423.1109P,2013Natur.493...62I,Pawlowski2020,2020MNRAS.499.3755S,2021arXiv210913253P}. Flattened distributions of galaxies are observed also on larger spatial scales out to $1-2$ Mpc in the Local Group \citep{Pawlowski2013} and in other nearby groups \citep{Muller2017,2020ApJ...891...18B,2021A&A...654A.161H,2021arXiv210608868M}. Thus, it is important to check the accuracy of these methods by comparing them with different approaches to estimate $M_{\rm dyn}$.

The Centaurus group is one of the best studied galaxy systems in our cosmic neighborhood \citep[e.g. ][]{1997AJ....114.1313C,2000AJ....119..593J,2002A&A...385...21K,2011A&A...530A..59C,2015A&A...583A..79M,2018ApJ...867L..15T}. Similarly to the Local Group, it is composed of two main giant galaxies -- Centaurus\,A (Cen\,A) and M83 -- each one with its own system of dwarf satellite galaxies. In this article we focus on Cen\,A and its dwarf galaxy satellites, which we will refer to as the ``Cen\,A group'' for simplicity. 

The combination of accurate distances based on the tip of the red giant branch (TRGB) method and line-of-sight velocity measurements for 28 dwarf galaxies show that the Cen A satellite system forms a
flattened and kinematically-coherent structure \citep{2015ApJ...802L..25T,Muller2016,Muller2018, Muller2021b}, analogous to those found around the Milky Way and M31. Satellite systems with a similar kinematic coherence are extremely rare in $\Lambda$CDM simulations, leading to the so-called ``planes of satellites'' problem \citep{2010A&A...523A..32K,2015MNRAS.452.1052L,Pawlowski2018}. These observations also suggest that dynamical mass estimates of Cen\,A should consider a flattened (non-spherical) system with both rotation and pressure support.

In this article, we use a Bayesian model to show that the satellite system of Cen\,A has significant rotational support (Sect.\,\ref{sec:kinematics}). Next, we use the Jeans' equation in cylindrical symmetry to estimate the circular velocity and dynamical mass of the Cen\,A group (Sect.\,\ref{sec:mass}). Finally, we compare our new mass estimate with previous determinations  in the literature and discuss the implications for $\Lambda$CDM cosmology and alternative theories (Sect.\,\ref{sec:disc}).

\section{Kinematics of Cen\,A satellite system}\label{sec:kinematics}

Cen\,A has 42 confirmed satellites and 30 additional candidates that await membership confirmation \citep{Muller2019}. Here we consider 27 galaxies from \citet{Muller2021b} that have both TRGB distances and line-of-sight velocities (see their Table A.1)\footnote{The references are:  \cite{1993AJ....105.1411P}, \cite{1999ApJ...524..612B},   \cite{2000AJ....119..593J,2000AJ....119..166J}, \citet{2004AJ....128...16K},  \cite{2005MNRAS.361...34D}, \cite{2007AJ....133.1756S}, \citet{2007AJ....133..261B},      \cite{2008ApJ...676..184T,2015ApJ...802L..25T}, \cite{2012MNRAS.420.2924K},        \cite{2008ApJ...674..909P}, \citet{2013AJ....145..101K}, \citet{Muller2019,Muller2021a}, and \cite{2020A&A...634A..53F}.}, as well as one dwarf galaxy with only velocity information, giving a total of 28 dwarf galaxies. For three galaxies the EDD team\footnote{Extragalactic Distance Database: http://edd.ifa.hawaii.edu/} \citep{2009AJ....138..323T} has re-reduced and updated the available HI data. The new values of the line-of-sight velocities are 468$\pm$2\,km s$^{-1}$ for NGC\,5102, 516$\pm$3\,km s$^{-1}$ for ESO\,324-024, and 545$\pm$2\,km s$^{-1}$ for ESO\,325-011, which we adopt here. {Throughout the paper, based on EDD we adopt a distance of 3.68\,Mpc for Cen\,A, we note however, that an average distance of 3.8$\pm$0.1\,Mpc for Cen\,A is found using different distance estimates \citep{2004A&A...413..903R,2010PASA...27..457H}. We use the EDD value to be as self-consistent as possible.} 

We build a Bayesian model to quantify the relative degrees of rotation and pressure support in the satellite system. Similar models have been used to study the motion of globular clusters within their host galaxies \citep[e.g. ][]{2001ApJ...559..828C,2016A&A...592A..55V,2020A&A...637A..26F,2020MNRAS.491L...1L}. 
Our model assumes that the satellite system is centered on Cen\,A and its galaxy members rotate in a common plane with a mean velocity $v_{\rm rot}$. Deviations from purely circular motions (out-of-planar motions and more complex orbits) are encapsulated in the parameter $\sigma_{\rm int}$ that is a proxy for the mean velocity dispersion of the satellite system. This parameter sums quadratically to the observational error $\delta_{v,k}$ on the line-of-sight velocity $v_{{\rm obs,} k}$ of a satellite $k$, giving the observed deviation from purely circular motions:
\begin{equation}
\sigma_{\rm obs, k}^2 =\sigma_{\rm int}^2+\delta_{v,k}^2.
\end{equation}

The projected velocity of a satellite galaxy along the line of sight is then given by
\begin{equation}\label{eq:vmod}
v_{\rm mod}(v_{\rm rot}, i, \rm{PA}) =v_{\rm rot} \sin{(\it{i})} \frac{-\Delta_{{\rm RA,} k} \sin{(\rm{PA})} + \Delta_{{\rm Dec,} k}\cos{(\rm{PA})}}{\it{D_{{\rm 3D,} k}}},
\end{equation}
where $i$ is the inclination of the plane with respect to the sky, PA is the position angle of the major axis of the plane, $D_{\rm 3D,k}$ is the measured 3D distance of a satellite $k$ from Cen\,A, and $\Delta_{{\rm RA,} k}$ and $\Delta_{{\rm Dec,}k}$ are its projected separations from Cen\,A in the right ascension and declination directions, respectively.

The log likelihood is then given by:
\begin{equation}
\log \mathcal{L}=\sum_{\rm k=1}^N{\log \left(\frac{1}{\sqrt{2\pi} \sigma_{\rm obs}}\right) -\frac{(v_{{\rm obs, k}}-(v_{\rm mod}+v_{\rm sys}))^2}{2\sigma_{\rm obs, k}^2}},
\end{equation}
where $N=28$ is the number of satellites and $v_{\rm sys}$ is the systemic velocity of the satellite system. Thus, the free parameters of the model are $v_{\rm rot}$, $\sigma_{\rm int}$, PA, $i$, and $v_{\rm sys}$.

For the systemic velocity, we assume a Gaussian prior centered at 556 km s$^{-1}$ (the mean of $v_{\rm obs, k}$) with a standard deviation of 10 km s$^{-1}$ (the standard error on the mean). The central value of $v_{\rm sys}$ is in close agreement with the systemic velocity of Cen\,A from HI observations \citep{2004AJ....128...16K}, planetary nebulae \citep{Peng2004b}, and globular clusters \citep{Peng2004a, Woodley2010}\footnote{However, velocities derived from stellar tracers are systematically lower (by 10-20 km s$^{-1}$) than the HI measurement. This is still within the error, but may indicate that the gas is moving in respect to the stellar component of Cen\,A, which could be explained by Cen\,A's recent major merger \citep{2020MNRAS.498.2766W}.}. For the inclination angle, we assume a Gaussian prior centered at 86$^{\circ}$ with a standard deviation of 1$^{\circ}$ as suggested by the 3D geometry of the satellite system \citep{Muller2019}. In general, the rotation velocity of a projected disk is degenerate with the inclination angle: a tight prior on $i$ is crucial to break this degeneracy. Given that the satellite plane is seen close to edge-on, the precise value of $i$ plays a minor role in our estimate of $v_{\rm rot}$ (see Eq. \ref{eq:vmod}).

\begin{figure}[ht]
    \centering
    \includegraphics[width=\linewidth]{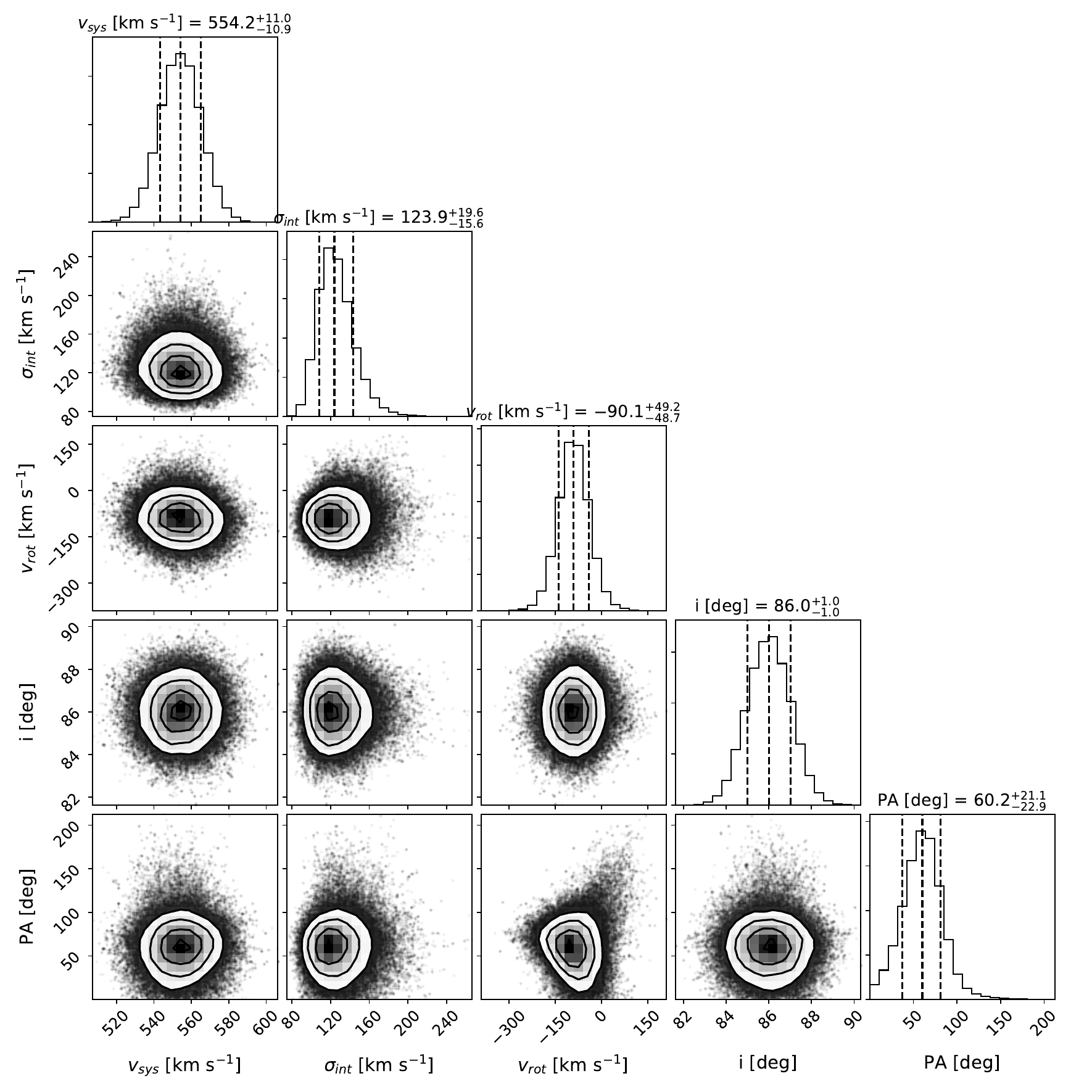}
    \caption{The posterior distribution of the five fitting parameters from the MCMC analysis of the Cen\,A satellite system. In the histograms, the three dashed lines indicate the 16, 50, and 84 percentiles, which correspond to the upper and lower uncertainty boundaries, and the best-fit parameter estimation (i.e. the median). See Sect.\,\ref{sec:kinematics} for details.}
    \label{fig:mcmc_disk}
\end{figure}
For the position angle, we run a preliminary MCMC simulation using a flat prior with $0^{\circ}<\textrm{PA}<360^{\circ}$. The resulting posterior probability distribution marginalized along PA displays two nearly Gaussian modes separated by $\sim$180$^{\circ}$, which represents two physical solutions for clockwise and anti-clockwise rotation. We shift one of the modes by 180$^{\circ}$ and determine PA$=63.7^{\circ} \pm 37.9^{\circ}$, which we use to set a Gaussian prior on PA in a final MCMC run. For $\sigma_{\rm int}$ and $v_{\rm rot}$, we use broad flat priors with $0<\sigma_{\rm int}<300$\,km s$^{-1}$ and $-400<v_{\rm rot}<400$\,km s$^{-1}$. To sample the posterior distribution, we run a MCMC with 100 walkers with a chain length of 10'000 each. We use 100 burn-in iterations and then sample over the full chains to get the posterior.

Figure\,\ref{fig:mcmc_disk} shows that the posterior distributions are well behaved and display distinct peaks, indicating that the maximum-likelihood parameters are well determined. We find a rotational component with $v_{\rm rot}=90\pm49$\,km s$^{-1}$ ($\sim$2$\sigma$ detection) and a random component with $\sigma_{\rm int}=124^{+20}_{-15}$\,km s$^{-1}$ ($\sim$7$\sigma$ detection). The resulting ratio $v_{\rm rot}/\sigma_{\rm int}$ is $\sim$0.7$^{+0.57}_{-0.42}$, indicating that rotational support is non-negligible. To test the significance of the rotational component, we re-run our experiment 100 times by reshuffling the velocities and their errors among the satellites (all other parameters are kept the same). This results in a lower rotational component ($9\pm47$\,km s$^{-1}$) and a higher random component ($130\pm5$\,km s$^{-1}$).
Most likely, the satellite system forms a thick structure flattened by rotation \citep[see also][]{Muller2019}. The final PA value of $60^{+21}_{-23}$ degrees is consistent with previous estimate within $\sim$2.5$\sigma$ \citep{Muller2018, Muller2021b}; the difference is due to the fact that the current measurement maximizes the rotation signal, while previous measurements maximized the number of kinematically coherent satellites considering only the sign of the line-of-sight velocities $v_{\rm obs, k}$ with respect to Cen\,A but not their absolute values. Our results for $\sigma_{\rm int}$ and $v_{\rm rot}$ are consistent with those from \citet{2006AJ....132.2424W}, who found $\sigma_{\rm int}=115\pm25$\,km s$^{-1}$, $v_{\rm rot}=125\pm50$\,km s$^{-1}$, and PA$=159\pm23^{\circ}$ using a spherical model for a smaller sample of 13 galaxies. The PA estimates differs significantly, though.

\begin{figure}[ht]
    \centering
    \includegraphics[width=\linewidth]{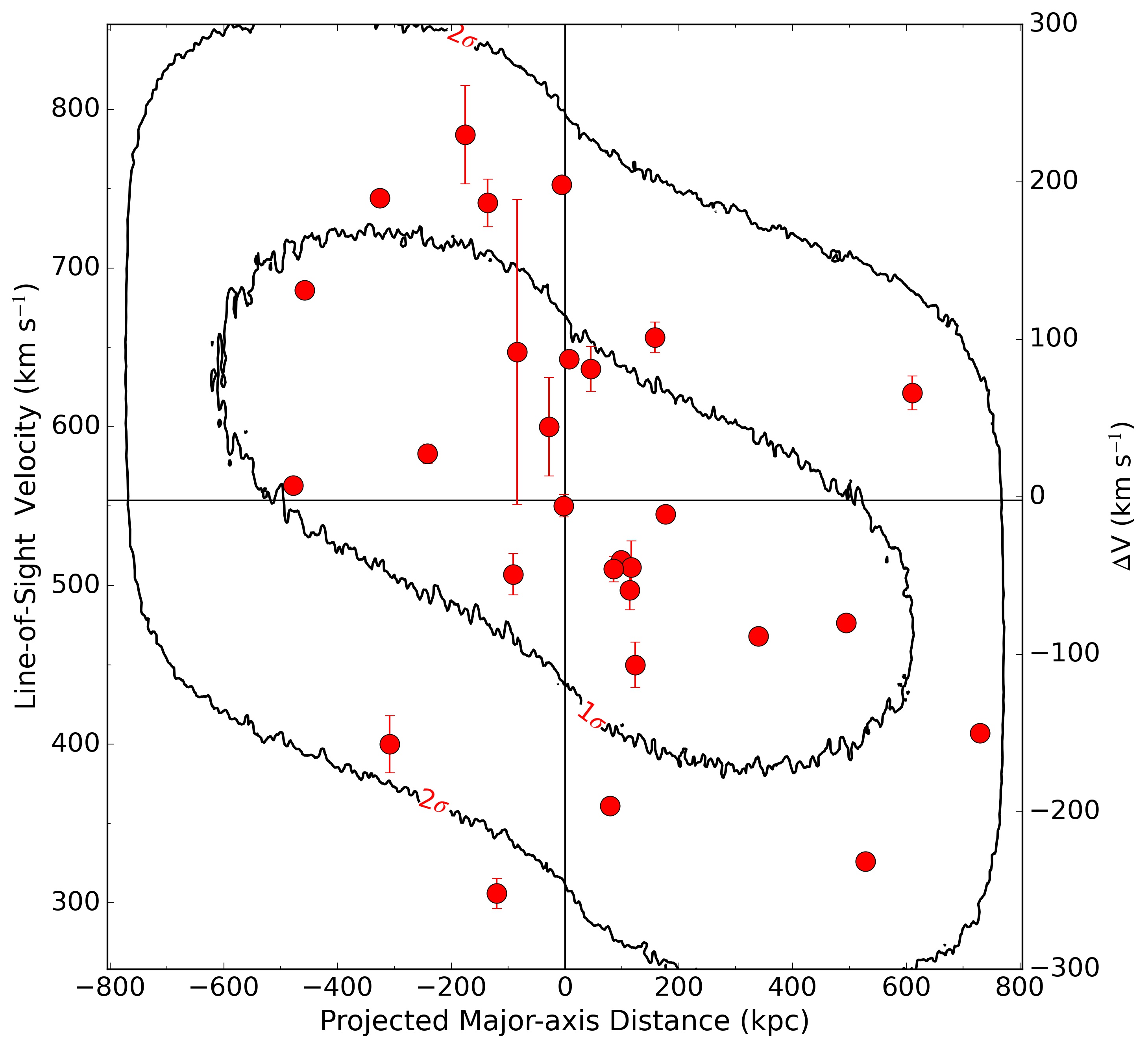}
    \caption{Position-Velocity diagram along the projected major axis of the satellite system. Red dots with 1$\sigma$ error bars show satellites of Cen\,A with measured line-of-sight velocities. The contours represent the best-fit rotating disk model with $v_{\rm rot}=90$ km\,s$^{-1}$ and $\sigma_{\rm int}=124$\,km s$^{-1}$. The inner and outer contours correspond to 1$\sigma$ and 2$\sigma$ probabilities of finding a galaxy at that location of the position-velocity diagram.}
    \label{fig:pvs}
\end{figure}
Figure\,\ref{fig:pvs} shows a position-velocity (PV) diagram along the projected major axis of the kinematic plane. The contours represent a rotating disk model projected on the sky, using the best-fit parameters from the MCMC analysis (Fig.\,\ref{fig:mcmc_disk}). The disk model has a Gaussian thickness of 130 kpc \citep{Muller2019} and assumes that satellite galaxies have equal probability to be found at any radius in the plane. Then, when the disk is projected to nearly edge-on orientation ($i=86^{\circ}$), the probability of having a galaxy at small projected distances is higher than having one at larger projected distances due to line-of-sight integration. Figure\,\ref{fig:pvs} shows that all galaxies but one -- KK\,203 -- agree within 2$\sigma$ with our model ($\sim$60$\%$ within 1$\sigma$). 

\section{Dynamics of Cen\,A satellite system}\label{sec:mass}

\subsection{Mean circular velocity}


We use the motion of satellite galaxies around Cen\,A to measure the circular velocity and dynamical mass of the whole group. A dynamical analysis is not straightforward because the satellite system is supported by both rotation and random motions {and only line-of-sight velocities are available}. We assume that the satellite system is axisymmetric and adopt a cylindrical reference system ($R$, $z$, $\phi$) where $z$ is perpendicular to the rotating plane defined by Eq.\,\ref{eq:vmod}. We further assume that the velocity ellipsoid is isotropic ($\sigma_{\rm R}^2 = \sigma_{\rm z}^2 = \sigma_{\phi}^2$), and thus the parameter $\sigma_{\rm int}$ in our MCMC analysis can be equate to the isotropic velocity dispersion. Under these simplifying assumptions, the circular velocity of a test particle subjected to the equilibrium gravitational potential is given by the Jeans' equation in cylindrical symmetry \citep[see, e.g.][]{2014A&A...563A..27L}:
\begin{equation}\label{eq:v}
v_{\rm circ}^2(R) = v_{\rm rot}^2(R)- A(R) \times \sigma_{\rm int}^2(R),
\end{equation}
where
\begin{equation}\label{eq:A}
   A(R)= \frac{\partial \ln \rho(R)}{\partial \ln R}+2\frac{\partial \ln \sigma_{\rm int}(R)}{\partial \ln R}
\end{equation}
depends on the density profile of the tracers $\rho(R)$ and the velocity dispersion profile $\sigma_{\rm int}(R)$. Having only 28 galaxies with line-of-sight velocities, we cannot constrain the velocity dispersion profile, so we consider $\sigma_{\rm int}$ as the average velocity dispersion across different radii and set the second term in Eq.\,\ref{eq:A} to zero. 

To estimate the first term in Eq.\,\ref{eq:A}, we consider all confirmed Cen\,A members adding 15 galaxies with known distances\footnote{These are: KKs53 \citep{2015ApJ...802L..25T},  CenA-MM-Dw11, CenA-MM-Dw5, CenA-MM-Dw4, CenA-MM-Dw10, CenA-MM-Dw6,  CenA-MM-Dw7, CenA-MM-Dw2, CenA-MM-Dw1, CenA-MM-Dw3, CenA-MM-Dw9, CenA-MM-Dw8 \citep{2014ApJ...795L..35C,2019ApJ...872...80C}, KK213, KK217, and CenN \citep{2002A&A...385...21K}.} (but without line-of-sight velocities) for a total of 43 objects. We project the galaxies in the plane ($R, \phi$) integrating along $z$, i.e. a face-on projection of the satellite system. We assume that the thickness of the plane does not change with radius, so $\partial \ln \rho/\partial \ln R = \partial \ln \Sigma / \partial \ln R$, where $\Sigma$ is the mass surface density. We measure $\Sigma(R)$ summing the baryonic mass of the galaxies in an inner circle of radius 200\,kpc and five outer annuli with a width of 150\,kpc. This gives five radial bins that contain 18, 8, 9, 5, and 3 galaxies from the innermost to the outermost bin. Stellar masses for the satellites are derived either from $K_s$-band magnitudes (where available) according to the Local Volume (LV) catalog adopting a mass-to-light ratio of 0.6 {(\citealt{2014AJ....148...77M,2015ApJ...802...18M}, the near infrared yields almost constant mass-to-light ratios)}, or from $V$-band magnitudes according to \citet{Muller2019} with a mass-to-light ratio of 2.0 \citep[see e.g.][for typical mass-to-light ratio values]{2017ApJ...836..152L}. When available, the cold gas mass is added to obtain the total baryonic mass, using $M_{\rm gas} =1.33 M_{\rm HI}$ where $M_{\rm HI}$ is the measured HI mass and the factor 1.33 takes the contribution of Helium into account. One galaxy (NGC4945) is removed from this analysis because it is two orders of magnitude more massive then the rest of the dwarf galaxies, leading to a sudden jump in the third radial bin.
Fitting the resulting density profile with a power law, we obtain $\partial \ln \Sigma / \partial \ln R =-3.8\pm1.5$ which is consistent with the outer slope of a NFW profile \citep{1997ApJ...490..493N}. {Because the bins were chosen arbitrarily and there are uncertainties in the stellar masses of the satellites, we double checked this fit using an MC approach. We randomly selected the  inner and  outer most ellipses to be within $\pm$100\,kpc of the initially chosen radii, and the radial bins having a width between 100\,kpc and 200\,kpc.  We varied the baryonic masses within a factor of 0.5 and 2. We repeatedly fitted the density profile using these randomizations and found that the slope of the density profile varied by $\pm$1.3, which is consistent with the error coming from the fit ($\pm$1.5). If we turn off the randomization of the mass-to-light ratios the variation doesn't change. }

We obtain a circular velocity of $v_{\rm circ}= 258\pm57$\,km s$^{-1}$. {The uncertainty of the circular velocity is derived through an error propagation of Eq.\,\ref{eq:v} using the rotational velocity $v_{rot}$, the velocity dispersion $\sigma_{int}$ and the fitted slope of the density profile $\partial \ln \Sigma / \partial \ln R$, contributing to 17\,km s$^{-1}$, 32\,km s$^{-1}$, and 44\,km s$^{-1}$, respectively.}

If we assume that there is some mild tangential anisotropy ($\sigma_{\rm R}^2 = \sigma_{\rm z}^2 = 2 \sigma_{\phi}^2$), Equation\,\ref{eq:A} is replaced by $\partial \ln \rho / \partial \ln R + 0.5$. Then, the circular velocity decreases by $\sim$15 km\,s$^{-1}$ which is smaller than our random errors on $v_{\rm circ}$ ($\sim$60 km\,s$^{-1}$).

\subsection{Dynamical and baryonic masses}

To estimate the dynamical mass of the group, we use the mean circular velocity and the maximal distance of the satellite population 
 \begin{equation}
M_{\rm tot}(<R_{\rm max})=\frac{R_{\rm max}v_{\rm circ}^2}{G}.
\end{equation}
With $v_{\rm circ}=258\pm57$\,kms $^{-1}$ and $R_{max}=801$\,kpc, we get a dynamical mass of $12.4\pm5.5\times10^{12}$\,M$_\odot$.

In a $\Lambda$CDM cosmological context, the properties of cosmic structures are usually given in terms of a density contrast $\Delta$ with respect to the critical density of the Universe $\rho_{\rm c}$. One then defines $R_\Delta$ as the radius at which the mass volume density is equal to $\Delta \rho_{\rm c}$. If we assume that $v_{\rm circ}$ corresponds to the circular velocity at radius $R_\Delta$, the total mass (baryons and dark matter) is given by \citep[see, e.g.][]{2012AJ....143...40M}:
 \begin{equation}
M_{\Delta}=(\Delta/2)^{-1/2}(GH_0)^{-1}v_{\rm circ}^3,
\end{equation}
where $H_0=75.1\pm3.8$\,km s$^{-1}$ Mpc$^{-1}$ \citep{2020AJ....160...71S} is the Hubble constant\footnote{{the exact value of the Hubble constant is highly debated, but generally ranges between 67 and 75\,km s$^{-1}$ Mpc$^{-1}$, with uncertainties of the order of 2-5 km s$^{-1}$ Mpc$^{-1}$ \citep{2019ApJ...882...34F,2020A&A...641A...6P,2020ApJ...902..145K,2021A&A...647A..72K,2021ApJ...908L...6R}. We here adopt a Hubble constant based on the baryonic Tully-Fisher relation (BTFR).}}, and $R_\Delta$ is given by
\begin{equation}
R_{\Delta}=\frac{G M_{\Delta}}{v_{\rm circ}^2}.
\end{equation}
Adopting $\Delta=200$, we derive $M_{200}= 5.3\pm3.5\times10^{12}$\,M$_\odot$ within $R_{200}=344$\,kpc.

{Now we need to estimate the baryonic mass of the Cen\,A group.}
This is the baryonic mass locked into all confirmed galaxy members, neglecting the possible inter-galactic medium and the gas mass of Cen\,A, which is negligible compared to its stellar mass ($M_{HI}/L_B = 0.01$, \citealt{2010A&A...515A..67S}). Cen\,A contributes for most of the group mass ($1.54\times10^{11}$\,M$_\odot$ from \citet{2012ApJS..203...17R}, re-scaled to a distance of 3.68\,Mpc), while the satellite galaxies add another $0.34\times10^{11}$\,M$_\odot$. 
{We estimate the baryonic mass of the Cen\,A group to be $1.9\times10^{11}$\,M$_\odot$.} We then derive a baryon fraction $f_{\rm b} = M_{\rm b}/M_{200} \simeq 0.035$. This estimate is significantly lower than the baryon fraction expected from $\Lambda$CDM fits to the cosmic microwave background: $\Omega_{\rm b}/\Omega_{\rm m} = 0.157$ \citep{Planck2020}. This is another facet of the so-called ``missing baryons problem'' \citep{2010ApJ...708L..14M}, which now emerges in a whole galaxy group rather than in a single galaxy. The amount of missing baryons in the Cen\,A group ($\sim$77$\%$) is comparable to that in typical massive galaxies \citep{2018MNRAS.480.4287K}. This discrepancy is usually explained assuming that the missing baryons reside in a hot, diffuse gas phase that is difficult to detect and quantify. In a $\Lambda$CDM context, therefore, we expect that the Cen\,A group should contain $\sim$8$\times10^{11}$ $M_{\odot}$ in hot gas. There is a hot X-ray halo and filamentary structure detected in association with Cen A \citep{1985ApJ...293..102F}.  Using most recent X-ray observations, \citet{Gaspari2019} find that the hot gas mass of Cen A is only $\sim 3 \times 10^8$ M$_\odot$ within 15 kpc (the observed size of the X-ray halo) and increases to $\sim 3 \times 10^{10}$ M$_\odot$ when extrapolating out to $\sim$300 kpc. This latter mass estimate is still $\sim$27 times smaller than that expected from the cosmic baryon fraction. 

Some authors use different density contrasts to define the characteristic quantities of a cosmic structure. If we use $\Delta=100$, we get $M_{100}=7.5\pm5.0\times10^{12}$\,M$_\odot$ within a virial radius $R_{100}=487$\,kpc. In this case, Cen\,A would have a baryon fraction $f_{\rm b} \simeq 0.025$ and be missing about 84$\%$ of the cosmologically available baryons. Clearly, the missing baryon problem in Cen\,A becomes even worse if we consider the dynamical mass of 1.2$\times$10$^{13}$\,M$_\odot$ out to the last measurable radius ($\sim$800 kpc): then the baryon fraction decreases to a mere $\sim$0.02 and the amount of missing baryons increases to $\sim$90$\%$.

\begin{table}[ht]
\caption{Dynamical masses determined for the Cen\,A group (excluding the M83 subgroup). Literature values are rescaled to a distance of 3.68 Mpc \citep{2015ApJ...802L..25T}.}
\begin{center}
\small
\setlength{\tabcolsep}{2pt} 
\begin{tabular}{l c c l}        
\hline\hline                 
Method & $M_{\rm tot}$ & Radius & Ref.\\
       & $10^{12}$\,M$_\odot$& kpc & \\
\hline      \\[-2mm]    
{Rotating Plane} & $2.0\pm0.4$ & $R_{1/2} = 130$ & 1\\\addlinespace[0.05cm]
{Virial Theorem}  & $1.6\pm0.4$ & $R_{1/2} = 130$ & 1\\\addlinespace[0.05cm]
Rotating Plane & $5.3\pm3.5$ & $R_{200} = 344$ & 1\\\addlinespace[0.05cm]
Virial Theorem &$\sim$8 &$\sim$400 & 2 \\\addlinespace[0.05cm] 
Rotating Plane & $7.5\pm5.0$ & $R_{100} = 487$ & 1 \\ \addlinespace[0.05cm] 
Virial Theorem & $\sim$12 & $\sim$600 & 3 \\\addlinespace[0.05cm] 
Rotating Sphere & $8.6\pm2.8$ & $\sim$760 & 4 \\\addlinespace[0.05cm] 
Rotating Plane & $12.4\pm5.5$ & $R_{\rm max}=800$ & 1 \\\addlinespace[0.05cm]
Zero-Velocity Surface & $\sim$6 &$\sim$1400 & 2  \\
\hline 
\end{tabular}
\end{center}
\footnotesize{References: (1) This work; (2) \citet{2007AJ....133..504K}; (3) \citet{2000AJ....119..609V}; (4) \citet{2006AJ....132.2424W}.}
\label{masses}
\end{table}

\section{Discussion}\label{sec:disc}

\subsection{Comparison with previous mass estimates}

The dynamical mass of Cen\,A has been estimated by several authors using different techniques. Studies based on globular clusters \citep{Peng2004a, Woodley2010} and/or planetary nebulae \citep{Peng2004b, Samurovic2016} probe the inner 40$-$80 kpc, so they trace the gravitational potential of the central galaxy. Here we focus on the dynamical mass of the whole Cen\,A group (excluding the M83 association). 

Table\,\ref{masses} lists mass measurements from the literature in increasing order of physical radius. Our mass estimate within $R_{\rm max}\simeq800$ kpc is consistent with previous measurements from \citet{2000AJ....119..609V} and \citet{2006AJ....132.2424W} within similar radii. Specifically, \citet{2000AJ....119..609V} used 30 candidate galaxy members (without secure distances at the time, {where later observations showed that some of their candidates were not Cen\,A members, but rather M\,83 members, which forms a distinct group, see \citealt{2015AJ....149...54T}}) assuming spherical symmetry and an isotropic velocity ellipsoid, whereas \citet{2006AJ....132.2424W} used 13 confirmed members assuming a spherical model with both rotation and dispersion support. We think that our modeling is more reliable because it considers the observed spatial flattening of the satellite systems and deprojects the rotation velocity from the sky plane to the disk plane (see Sect.\,\ref{sec:kinematics}).

Our mass estimate within $R_{\rm max}\simeq800$ kpc is significantly larger than the value from \cite{2007AJ....133..504K} at a larger radius ($\sim$1.4 Mpc) from the zero-velocity surface method. On the other hand, our mass estimates within a density contrast of 100 or 200, respectively ($R_{200}=344$\,kpc and $R_{100}=487$\,kpc), is consistent with the one from \cite{2007AJ....133..504K} within a similar radius ($\sim$400 kpc) using the Virial theorem.
We conclude that the zero-velocity method is inferior to estimates using spherical and/or disk models, which display only minor differences. 

To address the actual difference between a non-rotating spherical model {-- which can be considered as the $\Lambda$CDM expectation --} and a rotating disk model, we repeat the MCMC analysis in Sect. 2.2 dropping the $v_{\rm mod}$ term in Eq. (3) and fitting only for $\sigma_{\rm int}$, as it is routinely done in the study of pressure-supported systems (e.g. \citealt{1991AJ....102..914M,2009ApJ...704.1274W,2018A&A...618A.122T,2019ApJ...874L...5V,2019A&A...625A..76E,2021MNRAS.505.5686C}). We find $\sigma_{\rm int}=131 \pm 18$ km s$^{-1}$, which is larger than the previous estimate because the rotational component now enters as a pressure term. To infer the dynamical mass, we adopt the \citet{2010MNRAS.406.1220W} estimator that holds when the velocity dispersion profile is approximately flat near the half-mass radius $r_{1/2}$:
 \begin{equation}
M_{\rm tot}(<r_{\rm 1/2})= \frac{3 r_{\rm 1/2} \sigma_{\rm int}^2}{G},
\end{equation}
For $r_{1/2}=130$\,kpc (approximated by the median separation of the satellite system), we derive a dynamical mass of $1.6\pm0.4 \times 10^{12}$ M$_\odot$ for such a non-rotating spherical model. If we use Eq. 6 at the same radius, we derive a dynamical mass of $2.0\pm0.4 \times 10^{12}$ M$_\odot$  for the rotating disk model. These two values are consistent within the errors but we consider the rotating disk model better because the satellite system of Cen\,A is clearly non-spherical.

\subsection{Baryonic Tully-Fisher Relation and MOND}

\begin{figure}[ht]
    \centering
    \includegraphics[width=\linewidth]{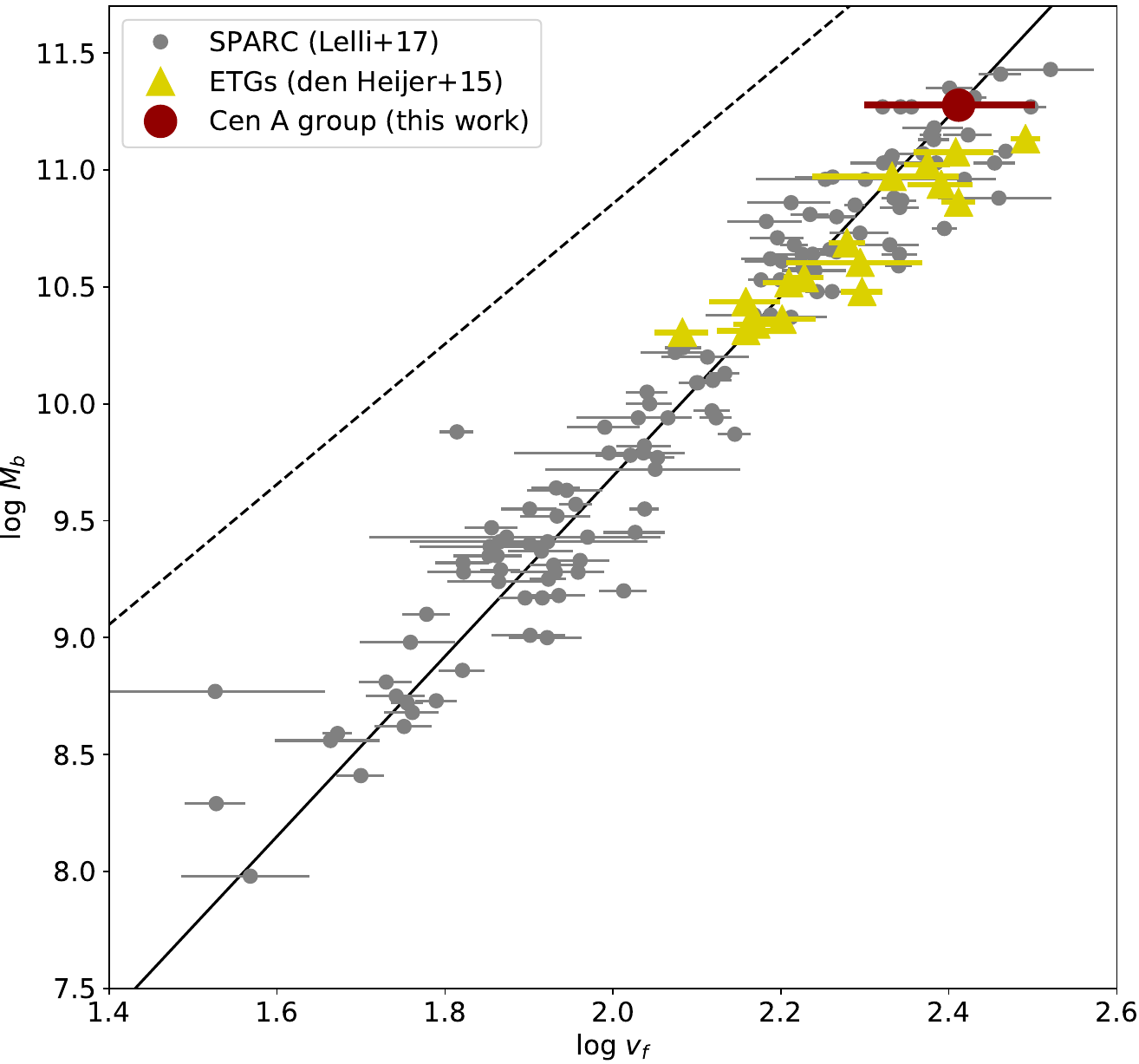}
    \caption{The BTFR of galaxies. The gray dots are late-type galaxies taken from the SPARC database \citep{2016ApJ...816L..14L,2019MNRAS.484.3267L}, the yellow triangles early-type galaxies (ETGs) from \citet{2015A&A...581A..98D}, and the red dot indicates our measurement of the Cen\,A group. The black line is the best fit of the BTFR, the dashed line the $\Lambda$CDM prediction assuming the  cosmic baryon fraction \citep{Planck2020}. \label{fig:BTFR}}
\end{figure}

Having measured the circular velocity and baryonic mass of the Cen\,A group, we can investigate its position on the baryonic Tully-Fisher relation (BTFR, \citealt{2000ApJ...533L..99M}). The BTFR is an empirical relationship between the baryonic mass of a galaxy and the circular velocity along the flat part of the rotation curve, pointing to a tight coupling between baryons and dark matter \citep[e.g.][]{2016ApJ...816L..14L, 2019MNRAS.484.3267L}. Figure\,\ref{fig:BTFR} shows the location of the Cen\,A group on the BTFR defined by late-type galaxies from the SPARC database \citep{SPARC} as well as early-type galaxies from Atlas$^{\rm 3D}$ \citep{2015A&A...581A..98D}. Our measurement of the Cen\,A group falls right on top of the BTFR. This is remarkable because we are comparing individual galaxies with a whole galaxy group, whose formation and evolution history is presumably governed by different processes on larger scales.

The location of Cen\,A on the BTFR agrees with the expectations of Milgromian Dynamics \citep[MOND,][]{Milgrom1983a, Milgrom1983b, Milgrom1983c}. MOND is an alternative to particle dark matter, in which the Newtonian laws of Gravity and/or inertia are modified at accelerations smaller than $\sim$10$^{-10}$ m\,s$^{-2}$ \citep{Famaey2012}. In a MOND context, the BTFR represents a fundamental law of Nature that should be followed by any \emph{isolated} gravitational system in equilibrium, independently of its formation and evolution. MOND has been successfully tested in Cen\,A using globular clusters \citep{Samurovic2016}. Fig.\,\ref{fig:BTFR} extends such a test out to much larger radii. In a MOND context, the Cen\,A group should contain little (if any) missing baryons. This is in line with the study of \citet{Milgrom2019} for a sample of 56 galaxy groups. Note that, in MOND, the internal gravity of the Cen\,A group would dominate over the external one within $\sim$500 kpc \citep{2021arXiv210910160O}, and a drop of ~60 km s$^{-1}$ would be expected at 800 kpc due to the external field effect arising from the cosmic large-scale structure (see e.g. \citealt{2019MNRAS.487.2441H,2021arXiv210904487F,2021ApJ...921..104C}), which is within our error bar for the global $v_{\rm circ}$ value. 

\section{Summary and Conclusions}

We studied the dynamics of the Cen\,A galaxy group using accurate 3D distances and line-of-sight velocities of its member galaxies. Our main results can be summarized as follows:
\begin{enumerate}
    \item We used a Bayesian model to study the kinematics of the satellite system considering both rotation and random motions. The ratio between the mean rotation velocity and the mean velocity dispersion is $\sim0.7$, indicating that the satellite system has significant rotational support.
    \item Assuming an axisymmetric and isotropic (in velocity space) satellite system, 
    we derived a mean circular velocity of 258 km s$^{-1}$. This translates into a dynamical mass of $1.2\times10^{13}$\,M$_\odot$ within 800\,kpc (the distance of the outermost satellite from Cen\,A). 
    \item In a $\Lambda$CDM context, we derive a virial mass $M_{200}=5.3\pm3.5\times10^{12}$\,M$_\odot$ within a virial radius $R_{200}=344$\,kpc. This gives a baryon fraction $M_{\rm b}/M_{200}=0.035$ implying that about 77$\%$ of the cosmologically available baryons are missing. The missing baryons may be in a hot diffuse medium with a mass of $\sim$8$\times$10$^{11}$ M$_\odot$, which is $\sim$4 times larger than the mass locked in stars and gas within galaxies. The expected hot gas mass is more than one order of magnitude larger than that inferred from the most recent X-ray observations. On the other hand, at the galaxy level, $M_*/M_{200} \approx 0.03$ for Cen\,A is much higher than predicted by common abundance matching relations (e.g., \citealt{2013MNRAS.428.3121M,2013ApJ...770...57B}) but is compatible within scatter and error bars with the stellar-to-halo-mass relation from \citet{2018AstL...44....8K}.
    \item Cen\,A group lies on the baryonic Tully-Fisher relation defined by individual galaxies. This is in agreement with MOND predictions with no need for a significant amount of unaccounted baryons.
\end{enumerate}

\begin{acknowledgements} 
{We thank the referee for the constructive report, which helped to clarify and improve the manuscript.}
O.M. is grateful to the Swiss National Science Foundation for financial support. O.M. also thanks the Arcetri Astrophysical Observatory for its hospitality during his visit. 
B.F., M.S.P. and O.M. thank the DAAD for PPP grant 57512596 funded by the BMBF, and the Partenariat Hubert Curien (PHC) for PROCOPE project 44677UE. B.F. acknowledges funding from the Agence Nationale de la Recherche (ANR projects ANR-18-CE31-0006 and ANR-19-CE31-0017), and from the European Research Council (ERC) under the European Union’s Horizon 2020 Framework programme (grant agreement number 834148).  M.S.P. is funded by a Leibniz-Junior Research Group grant (project number J94/2020) via the Leibniz Competition, and further thanks the Klaus Tschira Stiftung gGmbH and German Scholars Organization e.V. for support via a Klaus Tschira Boost Fund. K.F. acknowledges support through the European Space Agency fellowship programme.
\end{acknowledgements}

\bibliographystyle{aa}
\bibliography{aanda}
\end{document}